\documentclass[envcountsame]{llncs}

\usepackage{amsmath,amssymb}
\usepackage{graphicx}
\usepackage{wrapfig}
\usepackage{comment}

\newcommand{\eproof}{{\small\hspace{1mm}\raisebox{-1mm}{$\Box$}}}
 
\title{Fixed Point and Aperiodic Tilings}
\author{Bruno Durand\inst{1},
Andrei Romashchenko\inst{2,3}, Alexander Shen\inst{1,3}}

\institute{LIF, CNRS \& Univ. de Provence, Marseille
\and LIP, ENS de Lyon \& CNRS
\and Institute for Information Transmission Problems, Moscow}

\begin{document}

\pagestyle{plain}


\maketitle

\begin{abstract}

An aperiodic tile set was first constructed by R.~Berger while
proving the undecidability of the domino problem. It turned out
that aperiodic tile sets appear in many topics ranging from
logic (the Entscheidungsproblem) to physics (quasicrystals).

\smallskip

We present a new construction of an aperiodic tile set that is
based on Kleene's fixed-point construction instead of geometric
arguments. This construction is similar to J.~von Neumann
self-reproducing automata; similar ideas were also used by
P.~G\'acs in the context of error-correcting computations.

\smallskip

The flexibility of this construction allows us to construct a
``robust'' aperiodic tile set that does not have periodic (or
close to periodic) tilings even if we allow some (sparse enough)
tiling errors. This property was not known for any of the
existing aperiodic tile sets.

\end{abstract}

\section{Introduction}

In this paper\footnote{The first version of this preprint was published in {\tt arxiv} 
and {\tt hal}  on 18 Feb 2008. Later this paper was published in proceedings
of the DLT conference:  
\begin{quotation} 
\noindent
[DLT08] B.Durand, A.Romashchenko, A.Shen.  Fixed Point and Aperiodic Tilings. 
\emph{Proc. 12th international conference on Developments in Language Theory}. 
Kyoto, Japan, September 2008, pp. 537--548.
\end{quotation}
A short journal version of this work was presented in:
\begin{quotation} 
\noindent
[EATCS] B.Durand, A.Romashchenko, A.Shen. Fixed point theorem and aperiodic tilings. 
\emph{Bulletin of the EATCS \textup(The Logic in Computer Science Column by Yuri Gurevich\textup).} 
no~97 (2009) pp. 126--136.
\end{quotation}
Also this article   became a part of a long paper on a fixed-point technique in tilings:
\begin{quotation} 
\noindent
[DRS09] B.Durand, A.Romashchenko, A.Shen. Fixed-point tile sets and their applications. 2009, 
{\tt hal:00424024} and {\tt arxiv:0910.2415} (50 pages).
\end{quotation}
Since the present paper is only a preliminary preprint, 
we encourage the reader to refer directly to [EATCS] or [DRS09]
(this footnote is added on Jan 13, 2010).
}, 
\emph{tiles} are unit squares with colored sides.
Tiles are considered as prototypes: we may place translated
copies of the same tile into different cells of a cell paper
(rotations are not allowed). Tiles in the neighbor cells should
match (common side should have the same color in both).

Formally speaking, we consider a finite set $C$ of
\emph{colors}. A \emph{tile} is a quadruple of colors (left,
right, top and bottom ones), i.e., an element of $C^4$. A
\emph{tile set} is a subset $\tau\subset C^4$. A \emph{tiling}
of the plane with tiles from $\tau$ (\emph{$\tau$-tiling}) is a
mapping $U\colon \mathbb{Z}^2\to\tau$ that respects the color
matching condition. A tiling $U$ is \emph{periodic} if it has a
\emph{period}, i.e., a non-zero vector $T\in\mathbb{Z}^2$ such
that $U(x+T)=U(x)$ for all $x\in\mathbb{Z}^2$. Otherwise the
tiling is \emph{aperiodic}. The following classical result was
proved by Berger in a paper~\cite{berger} where he used this
construction as a main tool to prove \emph{Berger's theorem}:
the \emph{domino problem} (to find out whether a given tile set
has tilings or not) is undecidable.

\begin{theorem}
There exists a tile set $\tau$ such that $\tau$-tilings exist
and all of them are aperiodic.~\textup{\cite{berger}}
\end{theorem}

The first tile set of Berger
was rather complicated. Later many
other constructions were suggested. Some of them are simplified
versions of the Berger's construction (\cite{robinson}, see also
the expositions in
\cite{durand-gurevich,intelligencer,levin-arxiv}). Some others
are based on polygonal tilings (including famous Penrose and
Ammann tilings, see~\cite{grunbaum}). An ingenious construction
suggested in~\cite{kari} is based on the multiplication in a
kind of positional number system and gives a small aperiodic set
of $14$ tiles (in~\cite{culik} an improved version with $13$
tiles is presented). Another nice construction with a short
and simple proof (based explicitly on ideas of self-similarity)
was  recently  proposed by N.~Ollinger \cite{ollinger}.

In this paper we present yet another construction of aperiodic
tile set. It does not provide a small tile set; however, we find
it interesting because:

\textbullet\ The existence of an aperiodic tile set becomes a
simple application of a classical construction used in Kleene's
fixed point (recursion) theorem, in von Neumann's
self-reproducing automata~\cite{neumann} and, more recently, in
G\'acs' reliable cellular automata~\cite{gacs-focs,gacs}; we do
not use any geometric tricks. The construction of an aperiodic
tile set is not only an interesting result but an important tool
(recall that it was invented to prove that domino problem is
undecidable); our construction makes this tool easier to use
(see Theorem~\ref{substitution-tiling}).

\textbullet\ The construction is rather general, so it is
flexible enough to achieve some additional properties of the
tile set. Our main result is Theorem~\ref{main}: there exists a
``robust'' aperiodic tile set that does not have periodic (or
close to periodic) tilings even if we allow some (sparse enough)
tiling errors. It is not clear whether this can be achieved for
previously known aperiodic tile sets; however, the mathematical
model for a processes like quasicrystals' growth or
DNA-computation should take errors into account. Note that our
model (independent choice of place where errors are allowed) has
no direct physical meaning; it is just a simple mathematical
model that can be used as a playground to develop tools for
estimating the consequences of tiling errors.

The paper is organized as follows. In Section~\ref{macrotiles}
we define the notion of a self-similar tile set (a tile set that
simulates itself). In Section~\ref{simulating} we explain how a
tile set can be simulated by a computation implemented by
another tile set. Section~\ref{simulating-itself} shows how to
achieve a fixed point (a tile set that simulates itself). Then
we provide several applications of this construction: we use it
to implement substitution rules (Section~\ref{substitution}) and
to obtain tile sets that are aperiodic in a strong sense
(Section~\ref{strongly}) and robust to tiling errors
(Sections~\ref{filling} and~\ref{errors}). Section~\ref{islands}
provides probability estimates that show that tiling errors are
correctable with probability~$1$ (with respect to Bernoulli
distribution). Finally, we show some other applications of the
fixed point construction that simplify the proof of the
undecidability of the domino problem and related results.

\section{Macro-tiles}
        \label{macrotiles}
Fix a tile set $\tau$ and an integer $N>1$. A \emph{macro-tile}
is an $N\times N$ square tiled by matching $\tau$-tiles. Every
side of a macro-tile carries a sequence of $N$ colors called a
\emph{macro-color}.

Let $\rho$ be a set of $\tau$-macro-tiles. We say that $\tau$
\emph{simulates} $\rho$ if (a)~$\tau$-tilings exist, and (b)~for
every $\tau$-tiling there exists a unique grid of vertical and
horizontal lines that cuts this tiling into $N\times N$
macro-tiles from~$\rho$.

\begin{wrapfigure}[7]{r}{0pt}
\vbox{\vskip-9mm
\hbox{\includegraphics[scale=0.9]{fpte1.mps}}}
\caption{}
\label{fpt.1.mps}
\end{wrapfigure}

\textbf{Example 1}. Assume that we have only one (`white') color
and $\tau$ consists of a single tile with $4$ white sides. Fix
some $N$. There exists a single macro-tile of size $N\times N$.
Let $\rho$ be a singleton that contains this macro-tile. Then
every $\tau$-tiling can be cut into macro-tiles from $\rho$.
However, $\tau$ does not simulate $\rho$, since the placement
of cutting lines is not unique.

\textbf{Example 2}. In this example a set $\rho$ that consists
of exactly one macro-tile (that has the same macro-colors on all
four sides) is simulated by some tile set $\tau$. The tile set
$\tau$ consists of $N^2$ tiles indexed by pairs $(i,j)$ of
integers modulo~$N$. A tile from $\tau$ has colors on its sides
as shown on Fig.~\ref{fpt.1.mps}. The macro-tile in $\rho$ has
colors $(0,0),\ldots,(0,N-1)$ and $(0,0),\ldots,(N-1,0)$ on its
borders (Fig.~\ref{fpt.2.mps}).

\begin{wrapfigure}[11]{r}{0pt}
\vbox{\vskip-5mm%
\hbox{\includegraphics[scale=0.45]{fpte2.mps}}}
\caption{}
\label{fpt.2.mps}
\end{wrapfigure}

If a tile set $\tau$ simulates some set $\rho$ of
$\tau$-macro-tiles with zoom factor $N>1$ and $\rho$ is
isomorphic to $\tau$, the set $\tau$ is called
\emph{self-similar}. Here an \emph{isomorphism} between $\tau$
and $\rho$ is a bijection that respects the relations ``one tile
can be placed on the right of another one'' and ``one tile can
be placed on the top of another one''. (An isomorphism induces
two bijections between horizontal/vertical colors of $\tau$ and
horizontal/vertical macro-colors of $\rho$.)

The idea of self-similarity is used (more or less explicitly) in
most constructions of aperiodic tile sets (\cite{kari,culik} are
exceptions); we find the following explicit formulation useful.

\begin{theorem}
        \label{selfsimilar-aperiodic}
A self-similar tile set $\tau$ has only aperiodic tilings.
\end{theorem}

\textbf{Proof}.
Every $\tau$-tiling $U$ can be uniquely cut into $N\times
N$-macro-tiles from $\rho$. So every period $T$ of $U$ is a
multiple of $N$ (since the $T$-shift of a cut is also a cut).
Then $T/N$ is a period of $\rho$-tiling, which is isomorphic to
a $\tau$-tiling, so $T/N$ is again a multiple of $N$. Iterating
this argument, we conclude that $T$ is divisible by $N^k$ for
every $k$, so $T=0$.\eproof

So to prove the existence of aperiodic tile sets it is enough to
construct a self-similar tile set, and we construct it using the
fixed-point idea. To achieve this, we first explain how to
simulate a given tile set by embedding computations.

\section{Simulating a tile set}
        \label{simulating}
For brevity we say that a tile set $\tau$ simulates a tile set
$\rho$ when $\tau$ simulates some set of macro tiles $\tilde
\rho$ isomorphic to $\rho$ (e.g., a self-similar tile set
simulates itself).

Let us start with some informal discussion. Assume that we have
a tile set $\rho$ whose colors are $k$-bit strings
($C=\mathbb{B}^k$) and the set of tiles $\rho\subset C^4$ is
presented as a predicate $R(c_1,c_2,c_3,c_4)$. Assume that we
have some Turing machine $\mathcal{R}$ that computes $R$. Let us
show how to simulate $\rho$ using some other tile set $\tau$.

This construction extends Example~2, but simulates a tile set
$\rho$ that contains not a single tile but many tiles. We keep
the coordinate system modulo $N$ embedded into tiles of $\tau$;
these coordinates guarantee that all $\tau$-tilings can be
uniquely cut into blocks of size $N\times N$ and every tile
``knows'' its position in the block (as in Example~2). In
addition to the coordinate system, now each tile in $\tau$
carries supplementary colors (from a finite set specified below)
on its sides. On the border of a macro-tile (i.e., when one of
the coordinates is zero) only two supplementary colors (say, $0$
and $1$) are allowed. So the macro-color encodes a string of $N$
bits (where $N$ is the size of macro-tiles). We assume that
$N\ge k$ and let $k$ bits in the middle of macro-tile sides
represent colors from $C$. All other bits on the sides are zeros
(this is a restriction on tiles: each tile knows its coordinates
so it also knows whether non-zero supplementary colors are
allowed).

Now we need additional restrictions on tiles in $\tau$ that
guarantee that the macro-colors on sides of each macro-tile
satisfy the relation $R$. To achieve this, we ensure that bits
from the macro-tile sides are transferred to the central part of
the tile where the checking computation of $\mathcal{R}$ is
simulated (Fig.~\ref{fpt.3.mps}).

\begin{wrapfigure}[9]{l}{0pt}
\vbox{\vskip-6mm\hbox{%
\includegraphics[scale=0.75]{fpte3.mps}}}
\caption{}
\label{fpt.3.mps}
\end{wrapfigure}

For that we need to fix which tiles in a macro-tile form
``wires'' (this can be done in any reasonable way; let us assume
that wires do not cross each other) and then require that each
of these tiles carries equal bits on two sides; again it is easy
since each tile knows its coordinates.

Then we check $R$ by a local rule that guarantees that the
central part of a macro-tile represents a time-space diagram of
$\mathcal{R}$'s computation (the tape is horizontal, time goes
up). This is done in a standard way. We require that computation
terminates in an accepting state: if not, the tiling cannot be
formed.

To make this construction work, the size of macro-tile ($N$)
should be large enough: we need enough space for $k$ bits to
propagate and enough time and space (=height and width) for all
accepting computations of $\mathcal{R}$ to terminate.

In this construction the number of supplementary colors depends
on the machine $\mathcal{R}$ (the more states it has, the more
colors are needed in the computation zone). To avoid this
dependency, we replace $\mathcal{R}$ by a fixed universal Turing
machine~$\mathcal{U}$ that runs a \emph{program}
simulating~$\mathcal{R}$. Let us agree that the tape has an
additional read-only layer. Each cell carries a bit that is not
\begin{wrapfigure}[8]{r}{0pt}
\vbox{\vskip -6mm\hbox{%
\includegraphics[scale=0.7]{fpte4.mps}}}
\caption{}
\label{fpt.4.mps}
\end{wrapfigure}
changed during the computation; these bits are used as a program
for the universal machine (Fig.~\ref{fpt.4.mps}).
So in the computation zone the columns carry unchanged bits, and
the tile set restrictions guarantee that these bits form the
program for $\mathcal{U}$, and the central zone represents the
protocol of an accepting computation for that program. In this
way we get a tile set $\tau$ that simulates $\rho$ with zoom
factor $N$ using $O(N^2)$ tiles. (Again we need $N$ to be large
enough.)

\section{Simulating itself}
        \label{simulating-itself}
We know how to simulate a given tile set $\rho$ (represented as
a program for the universal TM) by another tile set $\tau$ with
a large enough zoom factor $N$. Now we want $\tau$ to be
isomorphic to $\rho$ (then Theorem~\ref{selfsimilar-aperiodic}
guarantees aperiodicity). For this we use a construction that
follows Kleene's recursion (fixed-point) theorem\footnote{%
    A reminder: Kleene's theorem says that for every
    transformation $\pi$ of programs one can find a program $p$
    such that $p$ and $\pi(p)$ produce the same output. Proof
    sketch: since the statement is language-independent (use
    translations in both directions before and after $\pi$), we
    may assume that the programming language has a function
    \texttt{GetText()} that returns the text of the program and
    a function \texttt{Exec(string s)} that replaces the current
    process by execution of a program \texttt{s}. (Think about an
    interpreter: surely it has an access to the program text; it 
    can also recursively call itself with another program.) Then 
    the fixed point is \texttt{Exec(}$\pi$\texttt{(GetText()))}.}
\cite{kleene}.

Note that most rules of $\tau$ do not depend on the program for
$\mathcal{R}$, dealing with information transfer along the
wires, the vertical propagation of unchanged program bits, and
the space-time diagram for the universal TM in the computation
zone. Making these rules a part of $\rho$'s definition (we
let $k=2\log N + O(1)$ and encode $O(N^2)$ colors by $2\log
N+O(1)$ bits), we get a program that checks that macro-tiles
behave like $\tau$-tiles in this respect.

The only remaining part of the rules for $\tau$ is the hardwired
program. We need to ensure that macro-tiles carry the same
program as $\tau$-tiles do. For that our program (for the
universal TM) needs to access the bits of its own text. (This
self-referential action is in fact quite legal: the program is
written on the tape, and the machine can read it.) The program
checks that if a macro-tile belongs to the first line of the
computation zone, this macro-tile carries the correct bit of the
program.

How should we choose $N$ (hardwired in the program)? We need it
to be large enough so the computation described (which deals
with $O(\log N)$ bits) can fit in the computation zone. The
computation is rather simple (polynomial in the input size,
i.e., $O(\log N)$), so for large $N$ it easily fits in
${\rm \Omega}(N)$ available time.

This finishes the construction of a self-similar aperiodic tile
set.

\section{Substitution system and tilings}
       \label{substitution}

The construction of self-similar tiling is rather flexible and
can be easily augmented to get a self-similar tiling with
additional properties. Our first illustration is the simulation
of substitution rules.

Let $A$ be some finite alphabet and $m>1$ be an integer. A
\emph{substitution rule} is a mapping $s\colon A\to A^{m\times
m}$. By $A$-configuration we mean an integer lattice filled with
letters from $A$, i.e., a mapping $\mathbb{Z}^2\to A$ considered
modulo translations.

A substitution rule $s$ applied to a configuration $X$ produces
another configuration $s(X)$ where each letter $a\in A$ is
replaced by an $m\times m$ matrix $s(a)$.

A configuration $X$ is \emph{compatible} with substitution rule
$s$ if there exists an infinite sequence
        $
\ldots \stackrel{s}{\to} X_{3}
\stackrel{s}{\to} X_{2}
\stackrel{s}{\to} X_{1}
\stackrel{s}{\to} X,
        $
where $X_i$ are some configurations.

\textbf{Example 3}. Let $A=\{0,1\}$, $s(0)=(\begin{smallmatrix}
0 & 1 \\ 1 & 0 \end{smallmatrix}),$ $s(1)=(\begin{smallmatrix} 0
& 1 \\ 1 & 0 \end{smallmatrix}).$ It is easy to see that the
only configuration compatible with~$s$ is the chess-board
coloring.

\textbf{Example 4}. Let $A=\{0,1\}$,
$s(0)=(\begin{smallmatrix} 0 & 1 \\ 1 & 0 \end{smallmatrix})$,
$s(1)=(\begin{smallmatrix} 1 & 0 \\ 0 & 1 \end{smallmatrix})$.
One can check that all configurations that are compatible with
this substitution rule (called \emph{Thue -- Morse
configurations} in the sequel) are aperiodic.

The following theorem goes back to~\cite{mozes}. It says that
every substitution rule can be enforced by a tile set.

\begin{theorem}[Mozes]
       \label{substitution-tiling}
Let $A$ be an alphabet and let $s$ be a substitution rule
over~$A$. Then there exists a tile set $\tau$ and a mapping
$e\colon\tau\to A$ such that

\textup{(a)}~$s$-image of any $\tau$-tiling is an $A$-configuration
compatible with $s$\textup;

\textup{(b)}~every $A$-configuration compatible with $s$ can be
obtained in this way.
\end{theorem}

\textbf{Proof}. We modify the construction of the tile set
$\tau$ (with zoom factor $N$) taking $s$ into account. Let us
first consider the very special case when

\textbullet\  the substitution rule maps each $A$-letter into an
$N\times N$-matrix (i.e., $m=N$).

\textbullet\ the substitution rule is easy to compute: given a
letter $u\in A$ and $(i,j)$, we can compute the $(i,j)$-th
letter of $s(u)$ in time ${\rm poly}(\log |A|)\ll N$.

In this case we proceed as follows. In our basic
construction every tile knows its
coordinates in the macro-tile and some additional information
needed to arrange `wires' and simulate calculations of the
universal TM. Now in addition to this basic structure each tile
keeps two letters of $A$: the first is the label of a tile
itself, and the second is the label of the $N\times N$-tile it
belongs to. This means that we keep additional $2\log |A|$ bits
in each tile, i.e., multiply the number of tiles by $|A|^2$. It
remains to explain how the local rules work. We add two
requirements:

(a) the second letter is the same for neighbor tiles (unless
they are separated by a border of some $N\times N$ macro-tile);

(b) the first letter in a tile is determined by the second
letter and the coordinates of the tile inside the macro-tile,
according to the substitution rule.

Both requirements are easy to integrate in our construction. The
requirement (a) is rather trivial; to achieve (b) we need to
embed in a macro-tile a calculation of $s([\mbox{label on this
macro-tile}])$. It is possible when $s$ is easy to compute.

The requirements (a) and (b) ensure that configuration is an
$s$-image of some other configuration. Also (due the
self-similarity) we have the same at the level of macro-tiles.
But this is not all: we need to guarantee that the first letter
on the level of macro-tiles is identical to the second letter on
the level of tiles. This is also achievable: the first letter of
a macro-tile is encoded by bits on its border, and we can
require that these bits match the second letter of the tiles at
that place (recall that second letter is the same across the
macro-tile). It is easy to see that now $\tau$ has the required
properties (each tiling projects into a configuration compatible
with~$\tau$ and vice versa).

However, this construction assumes that $N$ (the zoom factor) is
equal to the matrix size in the substitution rule, which is
usually not the case ($m$ is given, and $N$ we have to choose,
and it needs to be large enough). The solution is to let $N$ be
equal to $m^k$ for some $k$, and use the substitution rule
$s^k$, i.e., the $k$-th iteration of~$s$ (a configuration is
compatible with $s^k$ if and only if it is compatible with $s$).
Now we do not need $s$ to be easily computed: for large $k$ the
computation of $s^k$ will fit into the space available
(exponential in $k$). \eproof

\section{Strong version of aperiodicity}
   \label{strongly}

Let $\alpha>0$ be a real number. A configuration
$U\colon\mathbb{Z}^2\to A$ is \emph{$\alpha$-aperiodic} if for
every nonzero vector $T\in\mathbb{Z}^2$ there exists $N$ such
that in every square whose side is at least $N$ the fraction of
points $x$ such that $U(x)\ne U(x+T)$ exceeds~$\alpha$.

\textbf{Remark 1}. If $U$ is $\alpha$-aperiodic, then
Besicovitch distance between $U$ and any periodic pattern is at
least $\alpha/2$. (The Besicovitch distance is defined as
$\limsup_N d_N$ where $d_N$ is the fraction of points where two
patterns differ in the $N\times N$ centered square.)

\begin{theorem}
        \label{aperiodic}
There exists a tile set $\tau$ such that $\tau$-tilings exist
and every $\tau$-tiling is $\alpha$-aperiodic for every
$\alpha<1/3$.
\end{theorem}

\textbf{Proof}.
This tile set is obtained by applying
Theorem~\ref{substitution-tiling} to Thue--Morse substitution
rule $T$ (Example 4). Note that any configuration $C=\{c_{ij}\}$
compatible with $T$ is a xor-combination $c_{ij}=a_i\oplus b_j$
of two one-dimensional Thue-Morse sequences $a$ and $b$, and for
$a$ and $b$ a similar result (every shift changes between $1/3$
and $2/3$ of positions in a large block) is well
known (see, e.g.,~\cite{thue-morse-analysis}).\eproof

\section{Filling holes}
        \label{filling}
The second application of our flexible fixed-point construction
is an aperiodic
\begin{wrapfigure}[10]{r}{32mm}
\vbox{\vskip-0mm\hbox{%
\includegraphics[scale=0.75]{fpte5.mps}}}
\caption{}
\label{fpt.5.mps}
\end{wrapfigure}
tile set where isolated defects can be healed.

Let $c_1<c_2$ be positive integers. We say that a tile set
$\tau$ is \emph{$(c_1,c_2)$-ro\-bust} if the following holds: For
every $n$ and for every $\tau$-tiling $U$ of the
$c_2n$-neighborhood of a square $n\times n$ excluding the square
itself there exists a tiling $V$ of the entire
$c_2n$-neighborhood of the square (including the square itself)
that coincides with $U$ outside of the $c_1n$-neighborhood of
the square (see Fig.~\ref{fpt.5.mps}).

\begin{theorem}
        \label{robust-tileset}
There exists a self-similar tile set that is $(c_1,c_2)$-robust
for some $c_1$ and $c_2$.
\end{theorem}

\textbf{Proof}. For every tile set $\mu$ it is easy to
construct a ``robustified'' version $\mu'$ of $\mu$, i.e., a
tile set $\mu'$ and a mapping $\delta\colon\mu'\to\mu$ such
that: (a)~$\delta$-images of $\mu'$-tilings are exactly
$\mu$-tilings; (b)~$\mu'$ is ``5-robust'': every $\mu'$-tiling
of a $5\times 5$ square minus $3\times 3$ hole can be uniquely
extended to the tiling of the entire $5\times 5$ square.
 
\begin{wrapfigure}[6]{l}{13mm}
\vbox{\vskip-6.5mm\hbox{%
\includegraphics[scale=0.5]{fpte6.mps}}\vskip-1mm}
\caption{}
\label{fpt.6.mps}
\end{wrapfigure}

Indeed, it is enough to keep in one $\mu'$-tile the information
about, say, $5\times 5$ square in $\mu$-tiling and use the
colors on the borders to ensure that this information is
consistent in neighbor tiles.

This robustification can be easily combined with the fixed-point
construction. In this way we can get a $5$-robust self-similar
tile set $\tau$ if the zoom factors $N$ is large enough. Let us
show that this set is also $(c_1,c_2)$-robust for some $c_1$ and
$c_2$ (that depend on $N$, but $N$ is fixed.)

Indeed, let us have a tiling of a large enough neighborhood
around an $n\times n$ hole. Denote by $k$ the minimal integer
such that $N^k\ge n$ (so the $k$-level macro-tiles are greater
than the hole under consideration). Note that the size of
$k$-level macro-tiles is linear in $n$ since $N^k \le N\cdot n$.

In the tiling around the hole, an $N\times N$ block structure is
correct except for the $N$-neighborhood of the central $n\times
n$ hole. For similar reasons $N^2\times N^2$-structure is
correct except for the $N+N^2$-neighborhood, etc. So for the
chosen $k$ we get a $k$-level structure that is correct except
for (at most) $9=3\times 3$ squares of level $k$, and such a
hole can be filled (due to $5$-robustness) with $N^k\times N^k$
squares, and these squares can be then detalized back.

To implement this procedure (and fill the hole), we need a
correct tiling only in the $O(N^k)$-neighborhood of the hole
(technically, we need to have a correct tiling in 
$(3N^k)$-neighborhood of the hole; as $3N^k\le 3Nn$, we let $c_2=3N$).
The correction procedure involves  changes in another
$O(N^k)$-neighborhood of the hole (technically, changes  touch
$(2N^k)$-neighborhood of the hole; $2N^k\le 2Nn$, so we let
$c_1=2N$).
\eproof

\section{Tilings with errors}
        \label{errors}
Now we combine our tools to prove that there exists a tile set
$\tau$ that is aperiodic in rather strong sense: this set does
not have periodic tilings or tilings that are close to periodic.
Moreover, this remains true if we allow the tiling to have some
``sparse enough'' set of errors. Tiling with errors is no more a
tiling (as defined above): in some places the neighbor colors do
not match. Technically it is more convenient to consider
tilings with ``holes'' (where some cells are not tiled) instead
of errors but this does not matter: we can convert a tiling
error into a hole just by deleting one of two non-matching
tiles.

Let $\tau$ be a tile set and let $H\subset\mathbb{Z}^2$ be some
set ($H$ for ``holes''). We consider \emph{$(\tau,H)$-tilings},
i.e., mappings
        $
U\colon \mathbb{Z}^2\setminus H \to \tau
        $
such that every two neighbor tiles from $\mathbb{Z}^2\setminus
H$ match (i.e., have the same color on the common side).

We claim that there exists a tile set $\tau$ such that
(1)~$\tau$-tilings of the entire plane exist and (2)~for every
``sparse enough'' set $H$ every $(\tau,H)$-tiling is far from
every periodic mapping $\mathbb{Z}^2\to\tau$.

To make this claim true, we need a proper definition of a
``sparse'' set. The following trivial counterexample shows that a
requirement of small density is not enough for such a
definition: if $H$ is a grid made of vertical and horizontal
lines at large distance $N$, the density of $H$ is small but for
any $\tau$ there exist $(\tau,H)$-tilings with periods that are
multiples of~$N$.

The definition of sparsity we use (see below) is rather
technical; however, it guarantees that for small enough
$\varepsilon$ a random set where every point appears with
probability $\varepsilon$ independently of other points, is
sparse with probability $1$. More precisely, for every
$\varepsilon\in(0,1)$ consider a Bernoulli probability
distribution $B_\varepsilon$ on subsets of $\mathbb{Z}^2$ where
each point is included in the random subset with probability
$\varepsilon$ and different points are independent.

\begin{theorem}
        \label{main}
There exists a tile set $\tau$ with the following properties:
\textup{(1)}~$\tau$-tilings of $\mathbb{Z}^2$ exist;
\textup{(2)}~for all sufficiently small $\varepsilon$ for almost
every \textup(with respect to $B_\varepsilon$\textup) subset
$H\subset\mathbb{Z}^2$ every $(\tau,H)$-tiling is at least
$1/10$ Besicovitch-apart from every periodic mapping
$\mathbb{Z}^2\to\tau$.
\end{theorem}

\textbf{Remark 2}. Since the tiling contains holes, we need to
specify how we treat the holes when defining Besicovitch
distance. We do \emph{not} count points in $H$ as points where
two mappings differ; this makes our statement stronger.

\textbf{Remark 3}. The constant $1/10$ is not optimal and can be
improved by a more accurate estimate.

\textbf{Proof}. Consider a tile set $\tau$ such that (a)~all
$\tau$-tilings are $\alpha$-aperiodic for every $\alpha<1/3$;
(b)~$\tau$ is $(c_1,c_2)$-robust for some $c_1$ and $c_2$. Such
a tile set can be easily constructed by combining the arguments
used for Theorem~\ref{robust-tileset} and
Theorem~\ref{aperiodic}.

Then we show (this is the most technical part postponed until
Section~\ref{islands}) that for small $\varepsilon$ a
$B_\varepsilon$-random set $H$ with probability $1$ has the
following ``error-correction'' property: every $(\tau,H)$-tiling
is Besicovitch-close to some $\tau$-tiling of the entire plane.
The latter one is $\alpha$-aperiodic, therefore (if Besicovitch
distance is small compared to $\alpha$) the initial
$(\tau,H)$-tiling is far from any periodic mapping.

For simple tile sets that allow only periodic tilings this
error-correction property can be derived from basic results in
percolation theory (the complement of $H$ has large connected
component etc.) However, for aperiodic tile sets this argument
does not work and we need more complicated notion of ``sparse''
set based on ``islands of errors''. We employ the technique
suggested in \cite{gacs-focs} (see also applications of ``islands
of errors'' in \cite{gray}, \cite{durand-romash}).

\section{Islands of errors}
        \label{islands}

Let $E\subset\mathbb{Z}^2$ be a set of points; points in $E$ are
called \emph{dirty}; other points are \emph{clean}. Let
$\beta\ge\alpha>0$ be integers. A set $X\subset E$ is an
\emph{$(\alpha,\beta)$-island} in~$E$ if:

(1)~the diameter of $X$ does not exceed~$\alpha$;

(2)~in the $\beta$-neighborhood of $X$ there is no other points
from $E$.

(Diameter of a set is a maximal distance between its elements;
the distance $d$ is defined as the maximum of distances along
both coordinates; $\beta$-neighborhood of $X$ is a set of all
points $y$ such that $d(y,x)\le\beta$ for some $x\in X$.)

It is easy to see that two (different) islands are disjoint (and
the distance between their points is greater than $\beta$).

Let $(\alpha_1,\beta_1)$, $(\alpha_2,\beta_2)$,\ldots be a
sequence of pairs of integers and $\alpha_i\le \beta_i$ for all
$i$. Consider the iterative ``cleaning'' procedure. At the first
step we find all $(\alpha_1,\beta_1)$-islands (\emph{rank $1$
islands}) and remove all their elements from $E$ (thus getting a
smaller set $E_1$). Then we find all $(\alpha_2,\beta_2)$-islands
in~$E_1$ (\emph{rank $2$ islands}); removing them, we get
$E_2\subset E_1$, etc. Cleaning process is \emph{successful} if
every dirty point is removed at some stage.

At the $i$th step we also keep track of the
$\beta_i$-neighborhoods of islands deleted during this step. A
point $x\in\mathbb{Z}^2$ is \emph{affected} during a step $i$ if
$x$ belongs to one of these neighborhoods.

The set $E$ is called \emph{sparse} (for given sequence
$\alpha_i,\beta_i$) if the cleaning process is successful, and,
moreover, every point $x\in\mathbb{Z}^2$ is affected at finitely
many steps only (i.e., $x$ is far from islands of large ranks).

The values of $\alpha_i$ and $\beta_i$ should be chosen in such
a way that:

(1)~for sufficiently small $\varepsilon>0$ a
$B_\varepsilon$-random set is sparse with probability~$1$
(Lemma~1 below);

(2)~if a tile set $\tau$ is $(c_1,c_2)$-robust and $H$ is sparse,
then any $(\tau,H)$-tiling is Besicovitch close to some
$\tau$-tiling of the entire plane (Lemmas~2 and~3).

\textbf{Lemma 1}. Assume that
        $
8\sum_{k<n} \beta_k < \alpha_n \le \beta_n
        $
for every $n$ and
        $
\sum_i \frac{\log \beta_i}{2^i} <\infty.
        $
Then for all sufficiently small $\varepsilon>0$ a
$B_\varepsilon$-random set is sparse with probability~$1$.

\begin{wrapfigure}[11]{l}{0pt}
\vbox{\vskip -5mm\hbox{%
\includegraphics[scale=0.8]{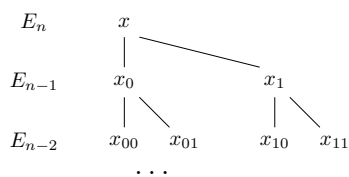}}}
\caption{Explanation tree; vertical lines connect different
names for the same points.}
\label{fpte7.mps}
\end{wrapfigure}
\textbf{Proof} of Lemma~1. Let us estimate the probability of the event
``$x$ is not cleaned after $n$ steps'' for a given point $x$
(this probability does not depend on $x$).
If $x\in E_n$, then $x$ belongs to $E_{n-1}$ and is not cleaned
during the $n$th step (when $(\alpha_n,\beta_n)$-islands in
$E_{n-1}$ are removed). Then $x\in E_{n-1}$ and, moreover, there
exists some other point $x_1\in E_{n-1}$ such that $d(x,x_1)$ is
greater than $\alpha_n/2$ but not greater than
$\beta_n+\alpha_n/2 < 2\beta_n$. Indeed, if there were no such
$x_1$ in $E_{n-1}$, then $\alpha_n/2$-neighborhood of $x$ in
$E_{n-1}$ is an $(\alpha_n,\beta_n)$-island in $E_{n-1}$ and $x$
would be removed.

Each of the points $x_1$ and $x$ (that we denote also $x_0$ to
make the notation uniform) belongs to $E_{n-1}$ because it
belongs to $E_{n-2}$ together with some other point (at the
distance greater than $\alpha_{n-1}/2$ but not exceeding
$\beta_{n-1}+\alpha_{n-1}/2$). In this way we get a tree
(Figure~\ref{fpte7.mps}) that explains why $x$ belongs to $E_n$.

The distance between $x_0$ and $x_1$ in this tree is at least
$\alpha_n/2$ while the diameter of the subtrees starting at
$x_0$ and $x_1$ does not exceed
        $
\sum_{i<n} 2\beta_i.
        $
Therefore, the Lemma's assumption guarantees that these subtrees
cannot intersect and, moreover, that all the leaves of the tree
are different. Note that all $2^n$ leaves of the tree belong to
$E=E_0$. As every point appears in $E$ independently from other
points, such an ``explanation tree'' is valid with probability
$\varepsilon^{2^n}$. It remains to estimate the number of possible
explanation trees for a given point~$x$.

To specify $x_1$ we need to specify horizontal and vertical
distance between $x_0$ and $x_1$. Both distances do not exceed
$2\beta_n$, therefore we need about $2\log (4\beta_n)$ bits to
specify them (including the sign bits). Then we need to specify
the distances between $x_{00}$ and $x_{01}$ as well as distances
between $x_{10}$ and $x_{11}$; this requires at most
$4\log(4\beta_{n-1})$ bits. To specify the entire tree we
therefore need
        $$
2\log (4\beta_n)+ 4 \log(4\beta_{n-1})+ 8 \log (4\beta_{n-2})+\ldots+
2^n \log (4\beta_1),
        $$
that is (reversing the sum and taking out the factor $2^n$) equal to
        $
2^n (\log (4\beta_1)+  \log(4\beta_2)/2+\ldots).
        $
Since the series $\sum \log \beta_n/2^n$ converges by
assumption, the total number of explanation trees for a given
point (and given $n$) does not exceed $2^{O(2^n)}$, so the
probability for a given point $x$ to be in $E_n$ for a
$B_\varepsilon$-random $E$ does not exceed
        $
\varepsilon^{2^n} 2^{O(2^n)},
        $
which tends to $0$ (even super-exponentially fast)
as~$n\to\infty$.

We conclude that the event ``$x$ is not cleaned'' (for a given
point $x$) has zero probability; the countable additivity
guarantees that with probability $1$ all points in
$\mathbb{Z}^2$ are cleaned.

It remains to show that every point with probability $1$ is
affected by finitely many steps only. Indeed, if $x$ is affected
by step $n$, then some point in its $\beta_n$-neighborhood
belongs to $E_n$, and the probability of
this event is at most
        $
O(\beta_n^2)\varepsilon^{2^n} 2^{O(2^n)}=
2^{2\log \beta_n +O(2^n)-\log(1/\varepsilon)2^n};
        $
the convergence conditions guarantees that $\log\beta_n=o(2^n)$,
so the first term is negligible compared to others, the
probability series converges and the Borel--Cantelli lemma
gives the desired result. \eproof

\smallskip

The following (almost evident) Lemma describes the error
correction process.

\textbf{Lemma 2}. Assume that a tile set $\tau$ is
$(c_1,c_2)$-robust, $\beta_k>4c_2\alpha_k$ for every $k$ and a
set $H\subset \mathbb{Z}^2$ is sparse (with respect to
$\alpha_i$, $\beta_i$). Then every $(\tau,H)$-tiling can be
transformed into a $\tau$-tiling of the entire plane by changing
it in the union of $2c_1\alpha_k$-neighborhoods of rank $k$
islands (for all islands of all ranks).

\textbf{Proof} of Lemma~2. Note that $\beta_k/2$-neighborhoods
of rank $k$ islands are disjoint and large enough to perform the
error correction of rank~$k$ islands, since
$\beta_k>4c_2\alpha_k$.\eproof

\medskip

It remains to estimate the Besicovitch size of the part of the
plane changed during error correction.

\textbf{Lemma 3}. The Besicovitch distance between the original
and corrected tilings (in Lemma 2) does not exceed $O(\sum_k
(\alpha_k/\beta_k)^2)$. 
(Note that the constant in $O$-notation depends on $c_1$.)

\textbf{Proof} of Lemma~3. We need to estimate the fraction of
changed points in large centered squares. By assumption, the
center is affected only by a finite number of islands. For every
larger rank $k$, the fraction of points affected at the stage $k$ in
\emph{any} centered square does not exceed
$O((\alpha_k/\beta_k)^2)$: if the square intersects with the changed
part, it includes a significant portion of the unchanged part.
For smaller ranks the same is true for \emph{all large enough}
squares that cover completely the island affecting the center
point).\eproof

It remains to chose $\alpha_k$ and $\beta_k$. We have to satisfy
all the inequalities in Lemmas~1--3 at the same time. To satisfy
Lemma~2 and Lemma~3, we may let $\beta_k = ck\alpha_k$ for large
enough $c$. To satisfy Lemma~1, we may let
$\alpha_{k+1}=8(\beta_1+\ldots+\beta_k)+1$. Then $\alpha_k$ and
$\beta_k$ grow faster that any geometric sequence (like
factorial multiplied by a geometric sequence), but still $\log
\beta_i$ is bounded by a polynomial in $i$ and the series in
Lemma~1 converges.

With these parameters (taking $c$ large enough) we may guarantee
that Besicovitch distance between the original $(\tau,H)$-tiling
and the corrected $\tau$-tiling does not exceed, say $1/100$.
Since the corrected tiling is $1/5$-aperiodic and
$1/10+2\cdot(1/100)<1/5$, we get the desired result
(Theorem~\ref{main}).\nobreak\eproof

\section{Other applications of fixed point self-similar tilings}
\label{variable-zoom}

The fixed point construction of aperiodic tile set is flexible
enough and can be used in other contexts. For example, the
``zoom factor'' $N$ can depend on the level $k$ (number of
grouping steps).
 %
For this each macro-tile should have $k$
encoded at its sides; this labeling should be consistent when
switching to the next level. For a tile of level $k$ its
coordinates inside a macro-tile are integers modulo $N_{k+1}$,
so in total $\log k +O(\log N_{k+1})$ bits are required and
$N_k$ steps should be enough to perform addition modulo
$N_{k+1}$. This means that $N_k$ should not increase too fast or
too slow (say, $N_k=\log k$ is too slow and $N_{k+1}=2^{N_k}$ is
too fast). Also we need to compute $N_k$ when $k$ is known, so
we assume that this can be done in polynomial time in the length
of $k$ (i.e., $\log k$). These restrictions still allow many
possibilities, say, $N_k=\sqrt{k}$, $N_k=k$, $N_k=2^{(2^k)}$,
$N_k=k!$ etc.

This ``self-similar'' structure with variable zoom factor can be
useful in some cases. Though it is not a self-similar according
to our definition, one can still easily prove that any tiling is
aperiodic. Note that now the computation time for the TM
simulated in the central part increases with level, and this can
be used for a simple proof of undecidability of domino problem
(in the standard proof~\cite{berger,durand-gurevich} one needs
to organize the ``computation zone'' with some simple geometric
tricks). With our new construction it is enough (for a given TM
$M$) to add in the program the parallel computation of $M$ on
the empty tape; if it terminates, this destroys the tiling.
 %
This construction can be used to replace the constant $1/10$ in
Theorem~\ref{main} by any number less that $1$;
to provide a new
proof for the results of~\cite{dls} (a tileset whose tilings
have maximal Kolmogorov complexity) and extend them to tilings
with sparse errors; 
it can be also used in some other
applications of tilings. 
Here is another application of this construction. 
We say that a tile
set $\tau$ is \emph{$m$-periodic} if $\tau$-tilings exist and
for each of them the set of periods is the set of \emph{all}
multiples of $m$ (this is equivalent to the fact that both
vectors $(0,m)$ and $(m,0)$ are periods). Let $E$ [resp. $O$] be
all $m$-periodic tile sets for all even $m$ [resp. odd $m$].

\begin{theorem}
        \label{reduction}
The sets $E$ and $O$ are inseparable enumerable sets.
\end{theorem}

\smallskip

\noindent
\emph{Acknowledgments.}
The authors thank the participants of the Kolmogorov seminar
in Moscow  (working on the RFBR projects 
05-01-02803 and  06-01-00122-a) 
for many fruitful discussions.

\thispagestyle{empty}

\end{document}